\documentclass[aps,pra,twocolumn,groupedaddress]{revtex4-1}
\usepackage{graphicx}
\usepackage{subfigure}
\usepackage[separate-uncertainty = true]{siunitx}
\DeclareSIUnit{\atoms}{atoms}

\begin{document}

% Use the \preprint command to place your local institutional report
% number in the upper righthand corner of the title page in preprint mode.
% Multiple \preprint commands are allowed.
% Use the 'preprintnumbers' class option to override journal defaults
% to display numbers if necessary
%\preprint{}

%Title of paper
\title{Evaporative cooling from an optical dipole trap in microgravity}

% repeat the \author .. \affiliation  etc. as needed
% \email, \thanks, \homepage, \altaffiliation all apply to the current
% author. Explanatory text should go in the []'s, actual e-mail
% address or url should go in the {}'s for \email and \homepage.
% Please use the appropriate macro foreach each type of information

% \affiliation command applies to all authors since the last
% \affiliation command. The \affiliation command should follow the
% other information
% \affiliation can be followed by \email, \homepage, \thanks as well.
\author{Christian Vogt}
\author{Marian Woltmann}
\author{Sven Herrmann}
%\email[]{Your e-mail address}
%\homepage[]{Your web page}
%\thanks{}
%\altaffiliation{}
\affiliation{ZARM, University of Bremen, Am Fallturm, D-28359 Bremen,
	Germany}
\author{Henning Albers}
\author{Dennis Schlippert}
\author{Ernst M. Rasel}
\affiliation{
Institute of Quantum Optics and QUEST-Leibniz Research School, Leibniz Universit\"at Hannover, Welfengarten 1, D-30167 Hannover, Germany}

\author{Claus L\"ammerzahl}
\affiliation{ZARM, University of Bremen, Am Fallturm, D-28359 Bremen,
	Germany}
\collaboration{PRIMUS}
%Collaboration name if desired (requires use of superscriptaddress
%option in \documentclass). \noaffiliation is required (may also be
%used with the \author command).
%\collaboration can be followed by \email, \homepage, \thanks as well.
%\collaboration{}
%\noaffiliation

\date{\today}

\begin{abstract}
In recent years, cold atoms could prove their scientific impact not only on ground but in microgravity environments such as the drop tower in Bremen~\cite{Rudolph2010,Zoest2007}, sounding rockets~\cite{Becker2018} and parabolic flights~\cite{Barrett2016}. We investigate the preparation of cold atoms in an optical dipole trap, with an emphasis on evaporative cooling under microgravity. Up to $ 1\times10^{6} $ rubidium-87 atoms were optically trapped from a temporarily dark magneto optical trap during free fall in the droptower in Bremen. The efficiency of evaporation is determined to be equal with and without the effect of gravity. This is confirmed using numerical simulations that prove the dimension of evaporation to be three-dimensional in both cases due to the anharmonicity of optical potentials. These findings pave the way towards various experiments on ultra-cold atoms under microgravity and support other existing experiments based on atom chips but with plans for additional optical dipole traps such as the upcoming follow-up missions to current~\cite{Elliott2018} and past~\cite{Becker2018} spaceborne experiments.
% \PACS{PACS code1 \and PACS code2 \and more}
% \subclass{MSC code1 \and MSC code2 \and more}
\end{abstract}

% insert suggested PACS numbers in braces on next line
% insert suggested keywords - APS authors don't need to do this
%\keywords{}

%\maketitle must follow title, authors, abstract, \pacs, and \keywords
\maketitle

% body of paper here - Use proper section commands
% References should be done using the~\cite, \ref, and \label commands
%Ultra Cold atoms...
\section{Introduction}
Atom interferometry is a precise quantum tool that will enhance a broad variety of measurements, ranging from large scale phenomena like gravitational wave detection~\cite{Graham2017, Canuel2018, Gao2018} to short scale Casimir-Polder forces and everything in between~\cite{Cronin2009}. Its sensitivity largely benefits from operation in microgravity, due to long free evolution times between laser pulses~\cite{Peters2001a}. These can only be realized with ultra cold atomic ensembles because of their low expansion rates. \newline 
The preparation of ultracold atoms generally follows the same path. Atoms are laser cooled in a magneto-optical trap before they are transferred into a purely optical or magnetic potential, where they are further cooled by evaporative cooling. In this relatively slow process the temperature is decreased at the cost of atom losses. Even though the creation of BECs could be demonstrated by laser cooling mechanisms lately~\cite{Stellmer2013, Hu2017a, Urvoy2019}, the lowest expansion rates realized~\cite{Hogan2015, Rudolph2016} are based on evaporative cooling~\cite{Masuhara1988} and delta-kick collimation~\cite{Ammann1997}. The former can either be implemented by the rf-knife method in magnetic traps or by lowering the optical potential confining an atomic ensemble. \newline
In spite of optical traps being a commonly used tool in cold atom experiments to trap, cool and manipulate atoms with low or vanishing magnetic susceptibility, to create quantum matter, to establish periodic crystals made out of light and to exploit Feshbach resonances, evaporation therein has never been realized in microgravity before. Achievements such as the first BEC in space~\cite{Becker2018} or the realization of atom interferometry in microgravity~\cite{Muntinga2013} were based on magnetic traps, implemented on atom chips~\cite{Reichel1999, Keil2016}. Thanks to the complementary advantages of dipole traps with respect to the manipulation with atom chips, we anticipate many applications for combing both approaches. \newline
In parallel with this work, Condon et al~\cite{Condon2019} demonstrated a dipole trap on an Einstein elevator. While we focus on the dipole trap behavior in microgravity, they prioritize a long investigation time for the cold atomic ensemble in weightlessness. Therefore, the preparation of cold atoms is mainly executed before the microgravity phase. \newline
Our demonstration of the trapping and evaporative cooling process in weightlessness is an important stepping stone to realize this kind of experiments in future space missions. \newline

\begin{figure*}[ht]
	\centering
	\includegraphics[width=0.7\linewidth]{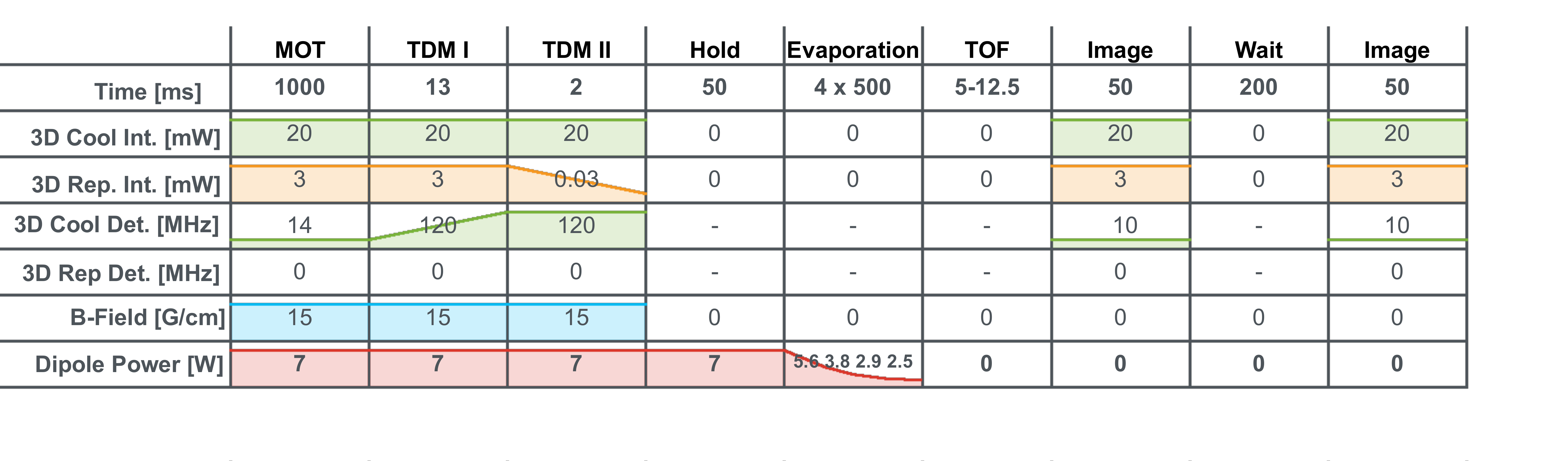}
	\caption{Depicted is the experimental sequence for the applied evaporation. The MOT loading is followed by a temporarily dark MOT scheme, the evaporation is interrupted at four different times. Fluorescence images where taken after varying time of flights to determine the ensembles temperature. The camera's shutter opens for \SI{600}{\micro\second} per image.}
	\label{fig:Sequence}     
\end{figure*}
\section{Experimental Setup}
\label{sec:Exp}
\begin{figure}
	\centering
	\includegraphics[width=0.3\textwidth]{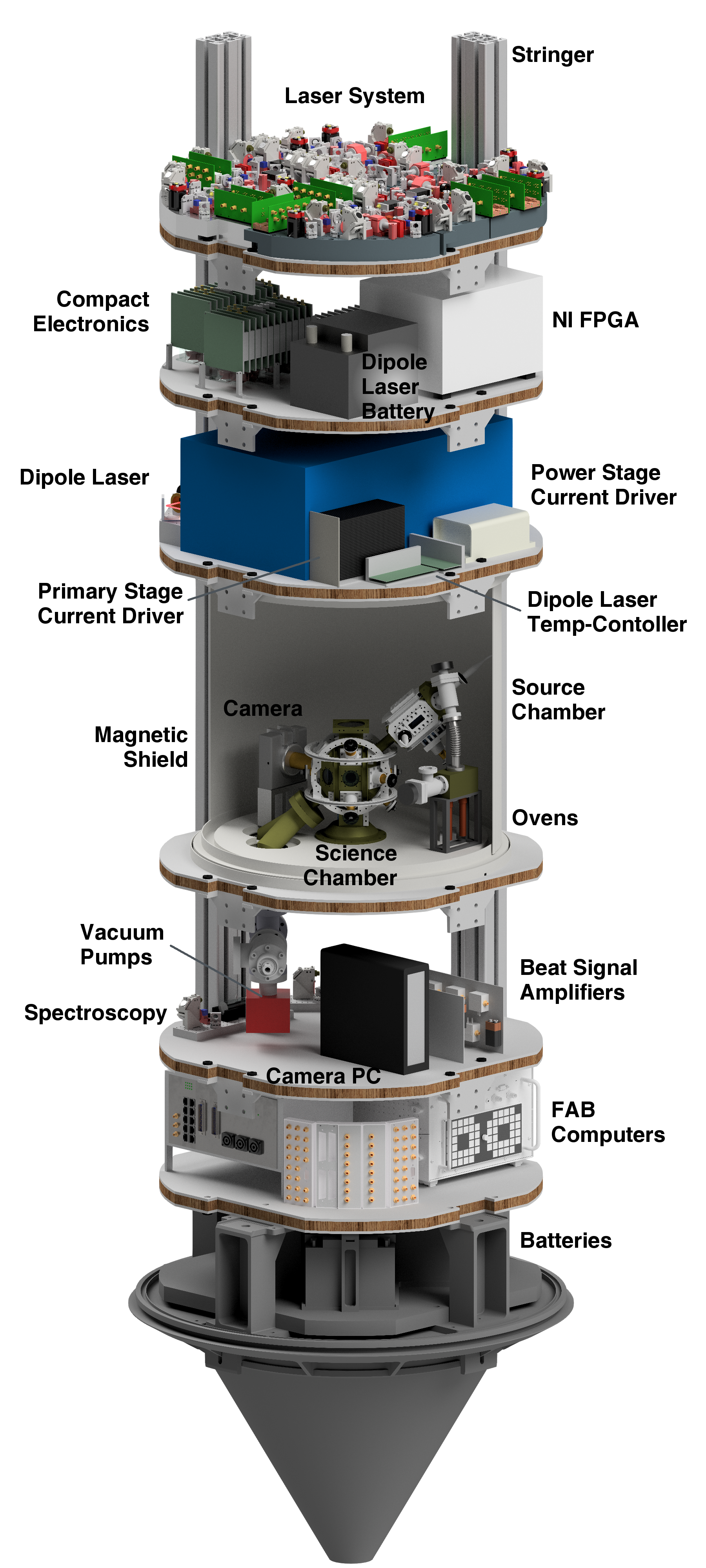}
	\caption{CAD drawing of the entire autonomous and compact experimental setup. The device is approximately \SI{2}{\meter} high with a diameter of \SI{70}{\centi\meter}.} 
	\label{fig:Setup}  
	\end{figure}
Our experimental setup (Figure~\ref{fig:Setup}) has been described in detail before~\cite{Kulas2017, Vogt2019}. Here we give a short summary of the main components and the techniques used in this work. The whole apparatus is portable and can be run autonomously with no supply connections. Since it is operated in the drop tower in Bremen, it fits into the standard drop tower capsule, which is \SI{2}{\meter} high and has a diameter of \SI{700}{\milli\meter}. The scientific payload is mounted on seven platforms with a total area of \SI{2.5}{\meter\squared} attached to a cage formed by four aluminum stringers. \newline
Atoms are released from an oven, a heated copper tube with a rubidium reservoir. To prevent molten rubidium or glass particles to drift into the vacuum chamber in microgravity, a bronze mesh is placed between oven and vacuum chamber. Atoms are precooled in a 2D$^+$MOT configuration~\cite{Schoser2002} with two retro-reflected beams and four magnetic coils in racetrack configuration. The atoms are optically pushed through a differential pumping tube into a 3D MOT consisting of six individual laser beams, which are not exactly power balanced due to a fixed fiber splitter. Cooling and repumping light are generated by semiconductor laser diodes with the cooling laser in MOPA (master oscillator power amplifier) configuration. \newline
The dipole trap is formed by a thulium fiber laser, emitting at \SI{1949}{\nano\meter}, with a maximum optical power of approximately \SI{7}{\watt} at the atoms' position. The beam intensity is regulated by a Pockels cell and two Glan-laser polarizers. The collimated beam is focused by a single lens down to a waist of \SI{45}{\micro\meter} into the center of the vacuum chamber. The vacuum quality of a few parts in \SI{E-9}{\hecto\pascal} is maintained by an ion-getter and two chemical pumps. The entire setup is capable of withstanding drops in the \SI{110}{\meter} drop tower with decelerations of up to 40 times Earth's gravitational acceleration. The available microgravity time is limited to \SI{4.7}{\second} per drop and can be used by our experiment twice a day. The performed experimental sequence is shown in Figure~\ref{fig:Sequence} and and allows for two datapoints per drop. \newline
% For one-column wide figures use
\begin{figure*}[htp]
	\centering
	\subfigure[]{\includegraphics[width=0.35\linewidth]{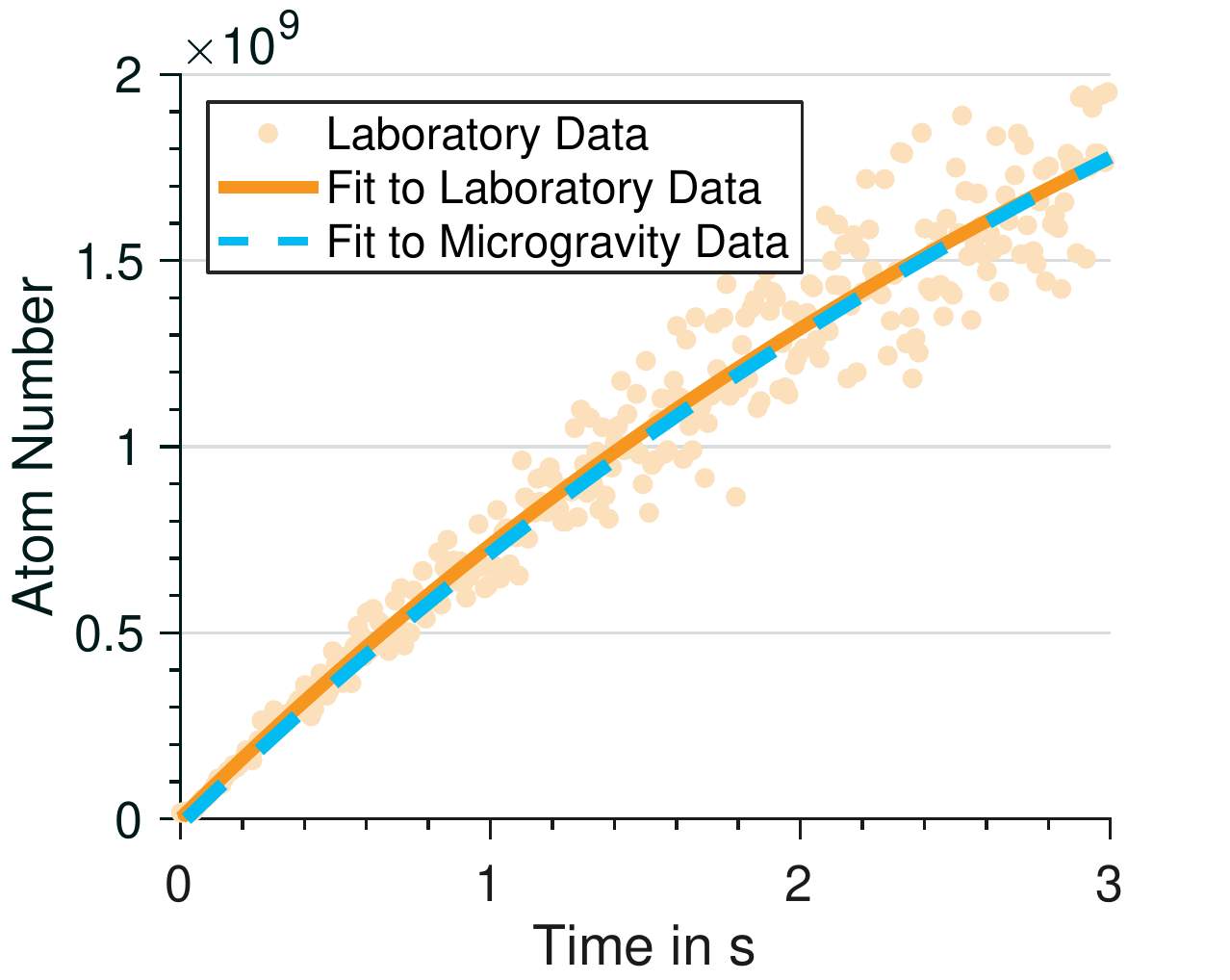}}\quad
	\subfigure[]{\includegraphics[width=0.35\linewidth]{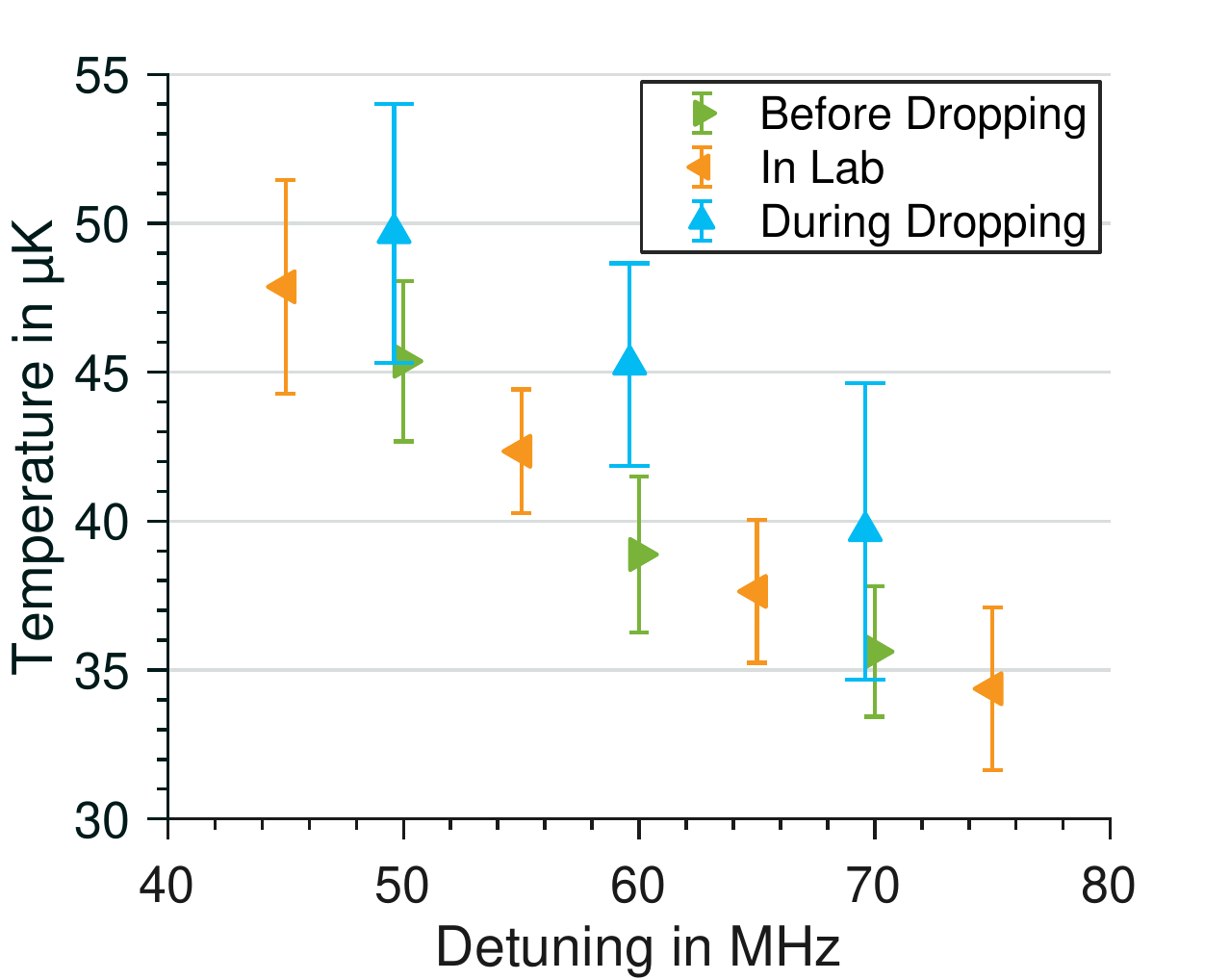}}	
	\caption{(a) Loading dynamics for our three-dimensional MOT in microgravity and on ground. Approximately \SI{1.5E9}{\atoms} are loaded in \SI{1.5}{\second}. The fluorescent signal for loading in gravity is plotted as orange dots and fitted to Equation \ref{Eq:MOT-Loadingrate} (orange line). For comparison, an equivalent dataset was recorded under microgravity and the resulting fit is shown as dashed blue line. For an improved visibility the underlying data points are omitted. (b) Comparison of the TDM performance in the laboratory (orange), in the droptower under gravity (green) and in microgravity (blue). In free fall the TDM efficiency is decreased leading to a temperature increase of a few \SI{}{\micro\kelvin}. This is attributed to less stable laser frequency locks.}
	\label{fig:MOT}
\end{figure*}
\section{Dipole Trap Loading}
\label{sec:Pre}
The precooling phase starts with a 3D-MOT which is fed by a  2D$^+$MOT. In this configuration $1\times10^9$ atoms of rubidium-87 can be trapped within one second, while the MOT performs equally well in microgravity and on ground (Figure~\ref{fig:MOT}(a)). The recorded atom number under gravity (orange dots) is fitted to 
\begin{equation}
N_{MOT}(t)=\frac{R_{0_{MOT}}}{\alpha}\left(1-e^{(-\alpha t)}\right),
\label{Eq:MOT-Loadingrate}
\end{equation}
where $N_{MOT}(t)$ is the number of atoms which are loaded into the MOT, $R_{0_{MOT}}$ is the initial trap loading rate, $\alpha$ describes losses caused by collisions with residual background gas atoms and $t$ is the loading time. The initial loading rate is determined to be \mbox{$R_{0_{MOT}}=$ \SI{8.32E8}{\atoms\per\second}} for operation with and without gravity. The same fit is displayed for a dataset recorded under microgravity as a dashed blue line.\newline
Since these atoms are about one order of magnitude too hot to be optically trapped in our setup, they have to be cooled further in order to be loaded into our dipole trap. The most efficient loading was obtained by employing a temporarily dark MOT (TDM) scheme in which the repumper intensity is decreased, while the cooling light is detuned further to the red with respect to the cooling transition. In addition to the cooling effect, this method compresses the atomic ensemble and thereby increases the transition efficiency into the optical potential. \newline
In the current setup the cooling light detuning is restricted to a maximum value of \SI{120}{\mega\hertz} with respect to the $F=2 \rightarrow F'=3$ transition. In combination with a decreased rempumping beam power of \SI{0.2}{\watt\per\square\meter} this leads to a temperature of \SI{28.0(5)}{\micro\kelvin}. In contrast to the MOT, the TDM performance is worsened in microgravity, as can be seen in Figure~\ref{fig:MOT}(b). This is attributed to less stable frequency locks during the drop, caused by sudden mechanical and electrical changes when releasing the capsule. Nevertheless the performance was found to be sufficient to effectively load an optical dipole trap.\newline   
\begin{figure}[bh]
	\centering
	% Use the relevant command to insert your figure file.
	% For example, with the graphicx package use
	\includegraphics[width=0.4\textwidth]{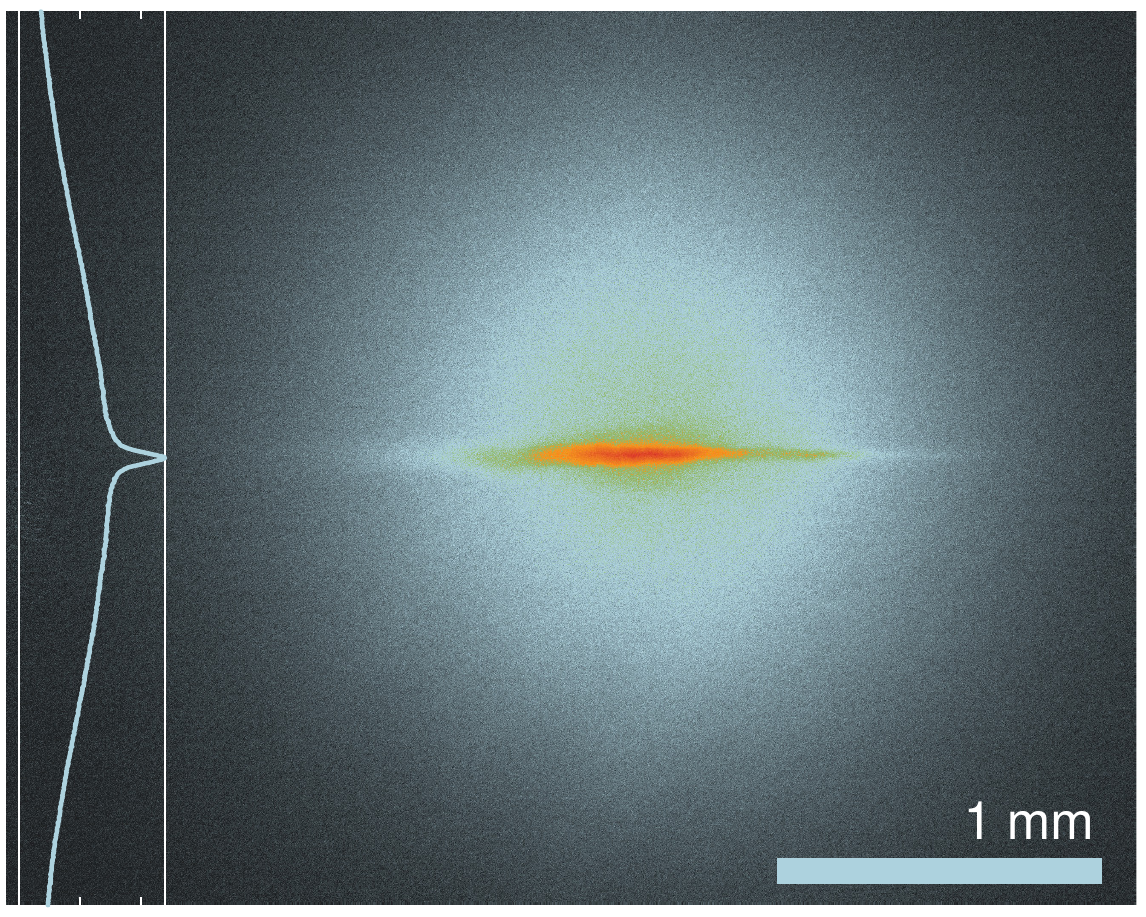}
	% figure caption is below the figure
	\caption{False colour image of the fluorescence of approximately 1 million rubidium atoms caught in an optical dipole trap in microgravity. The sharp structure in the middle represents trapped atoms while one can see non trapped atoms disappear radially. The picture was taken \SI{50}{\milli\second} after switching of the MOT light and magnetic field. On ground would have dropped out of the imaged area due to gravitational acceleration. The inset on the left side shows the summed fluorescent signals for every line in arbitrary units.}
	\label{fig:DipoleTrap}       
\end{figure}
Due to the choice of a dipole trapping laser with \SI{1949}{\nano\meter} wavelength, we are able to load the dipole trap directly from the MOT~\cite{Zaiser11PRA}, without an intermediate magnetic trapping or gray molasses scheme~\cite{Rosi2018}. This is enabled by the same, wavelength dependent sign of the complex polarisability for the ground- and excited state of the MOT transition in rubidium-87~\cite{Cornelussen2004}. Optimal loading was observed with the laser parameters determined in the former section but timing was found to be crucial. By splitting the TDM scheme into two separate steps for detuning and intensity reduction with durations of \SI{13}{\milli\second} and \SI{2}{\milli\second} respectively, $5\times 10^{6}$ atoms could be transferred into the optical trap. Furthermore, we can confirm a strong atom number dependency on the overlap between both traps~\cite{Kuppens2000}.\newline
The free beam path of the dipole trapping laser spanning more than \SI{1}{\meter} including mirrors on different platforms gave rise to pointing variations between drops. Therefore, this setup is not capable of precisely determining the loading characteristics in microgravity without further pointing stability improvements. In the recorded atom numbers, no significant difference between ground based and drop tower operation could be observed. \newline
Figure~\ref{fig:DipoleTrap} shows a fluorescent image of an optical dipole trap in microgravity. While one can see the residual TDM atoms disappear radially without the gravitational force, a bright line remains in the image center, representing the optically trapped atoms.\newline
\begin{figure*}[tp]
	\centering
	\subfigure[]{\includegraphics[width=0.35\linewidth]{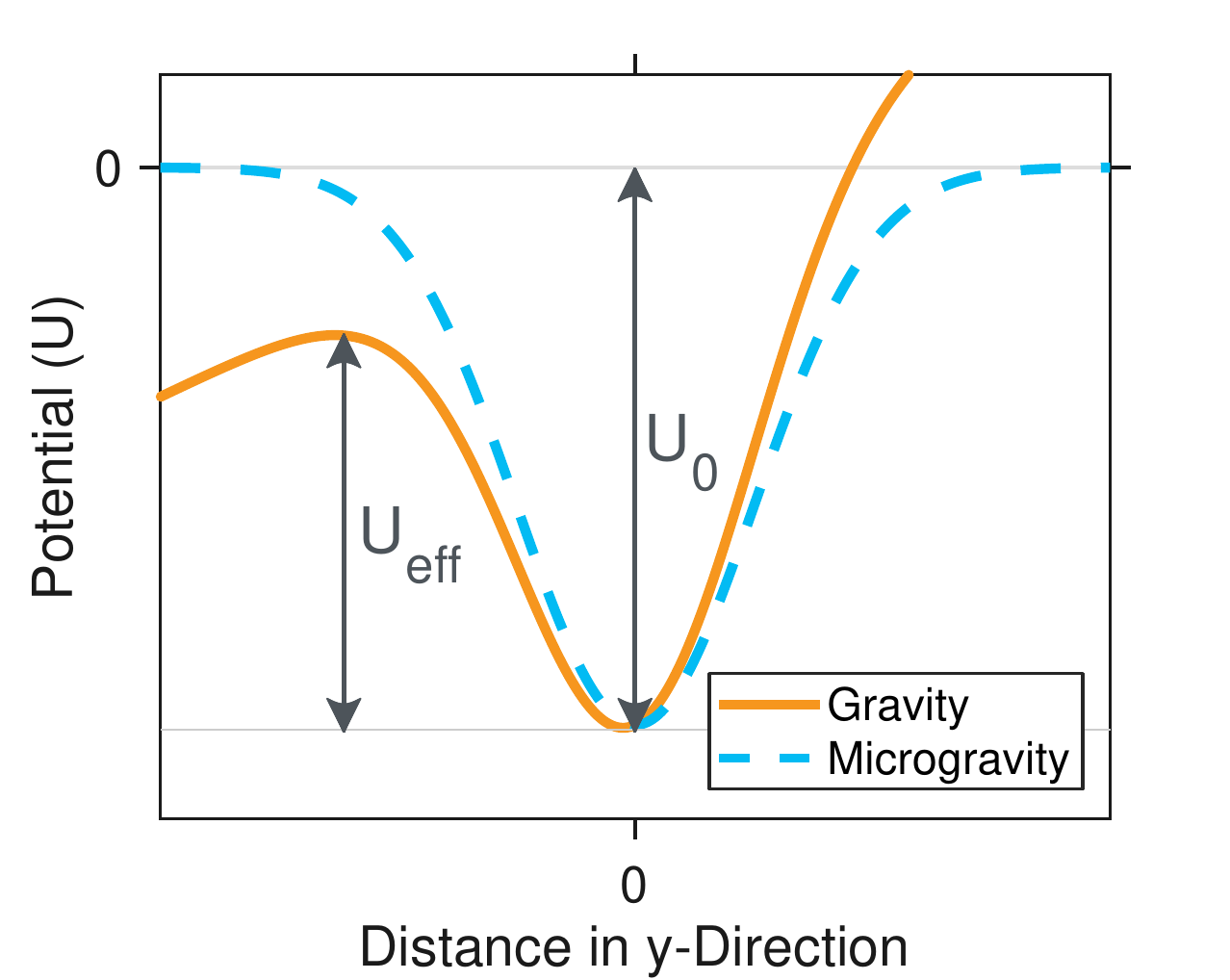}}
	\subfigure[]{\includegraphics[width=0.35\linewidth]{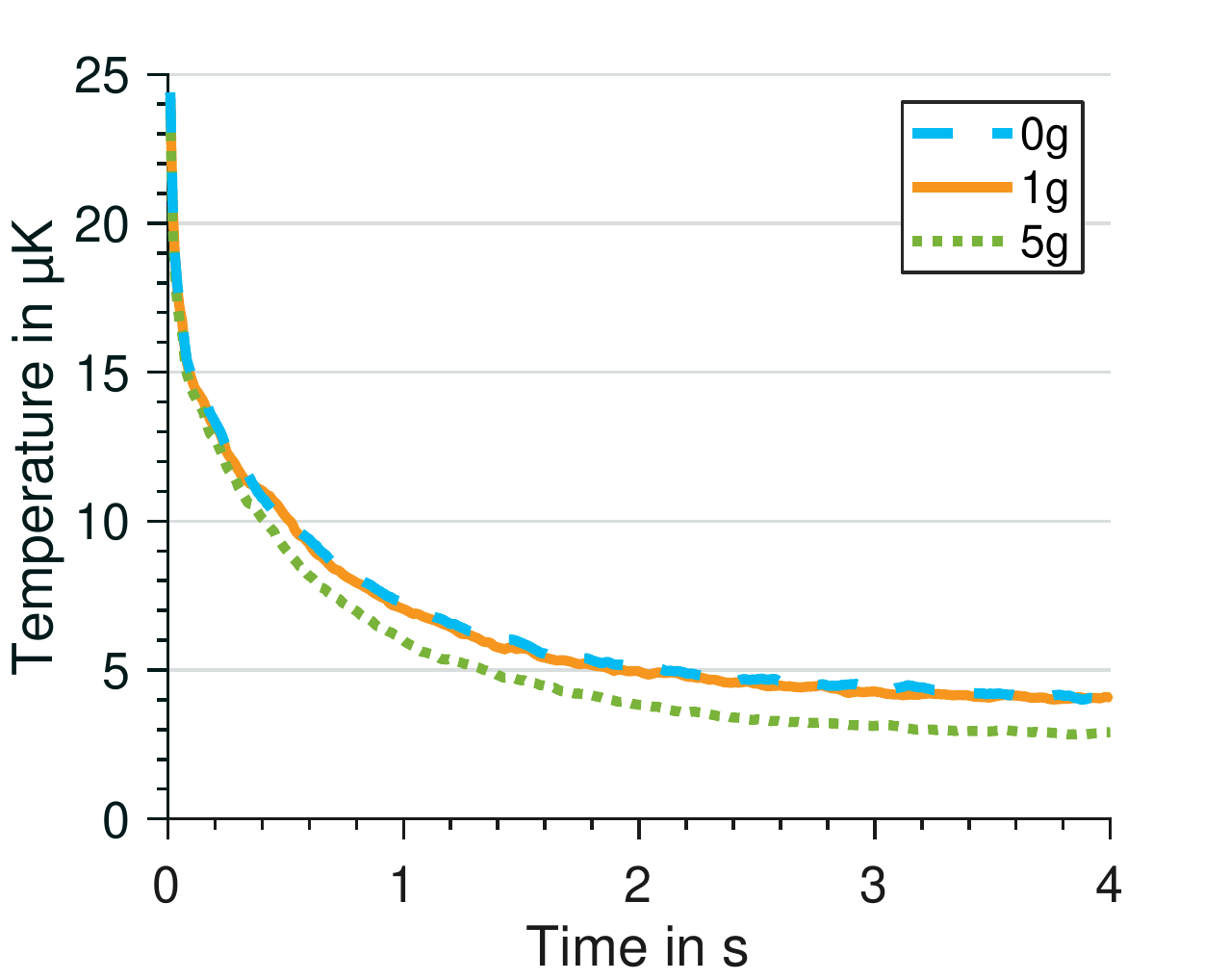}}	
	\subfigure[]{\includegraphics[width=0.35\linewidth]{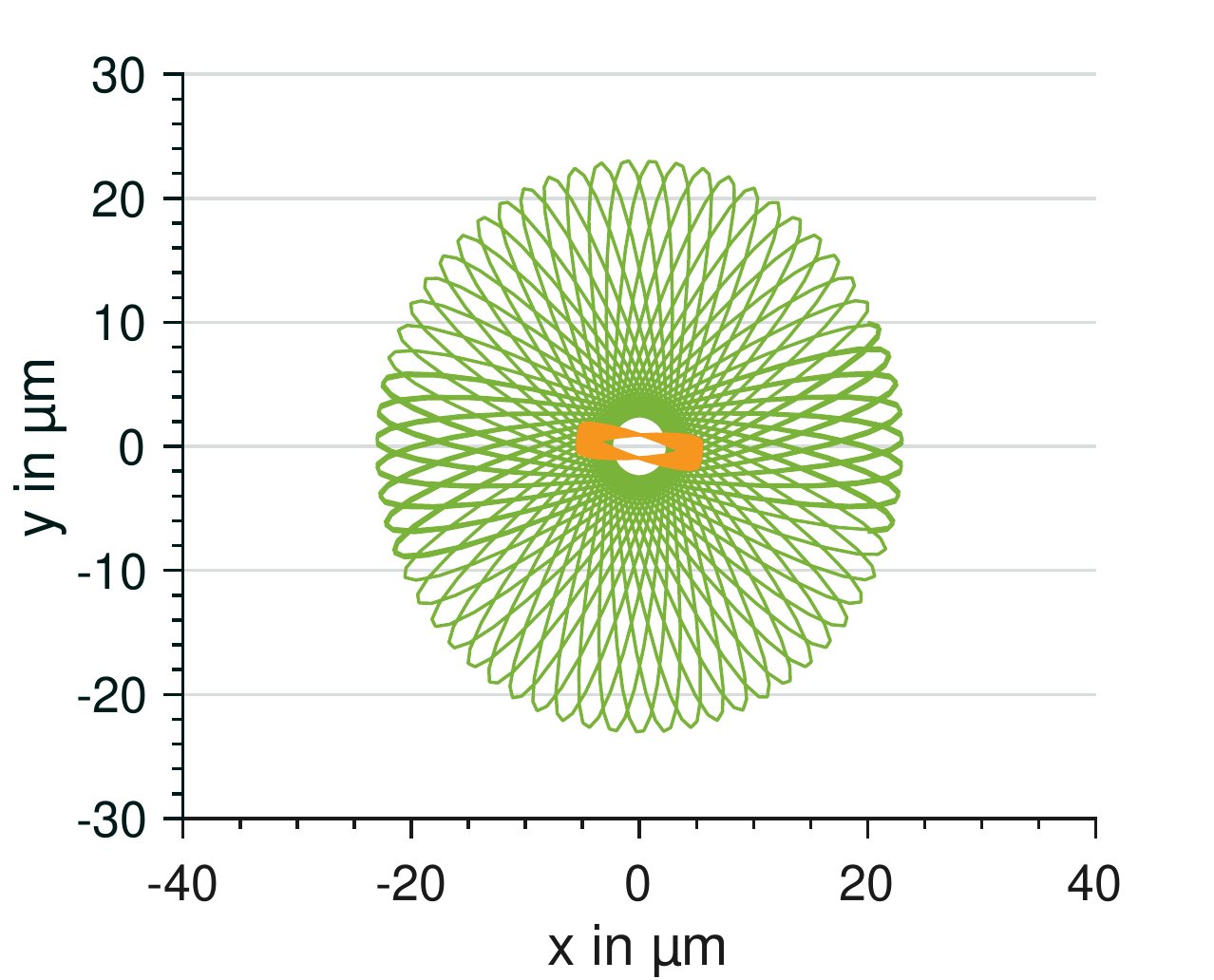}}
	\subfigure[]{\includegraphics[width=0.35\linewidth]{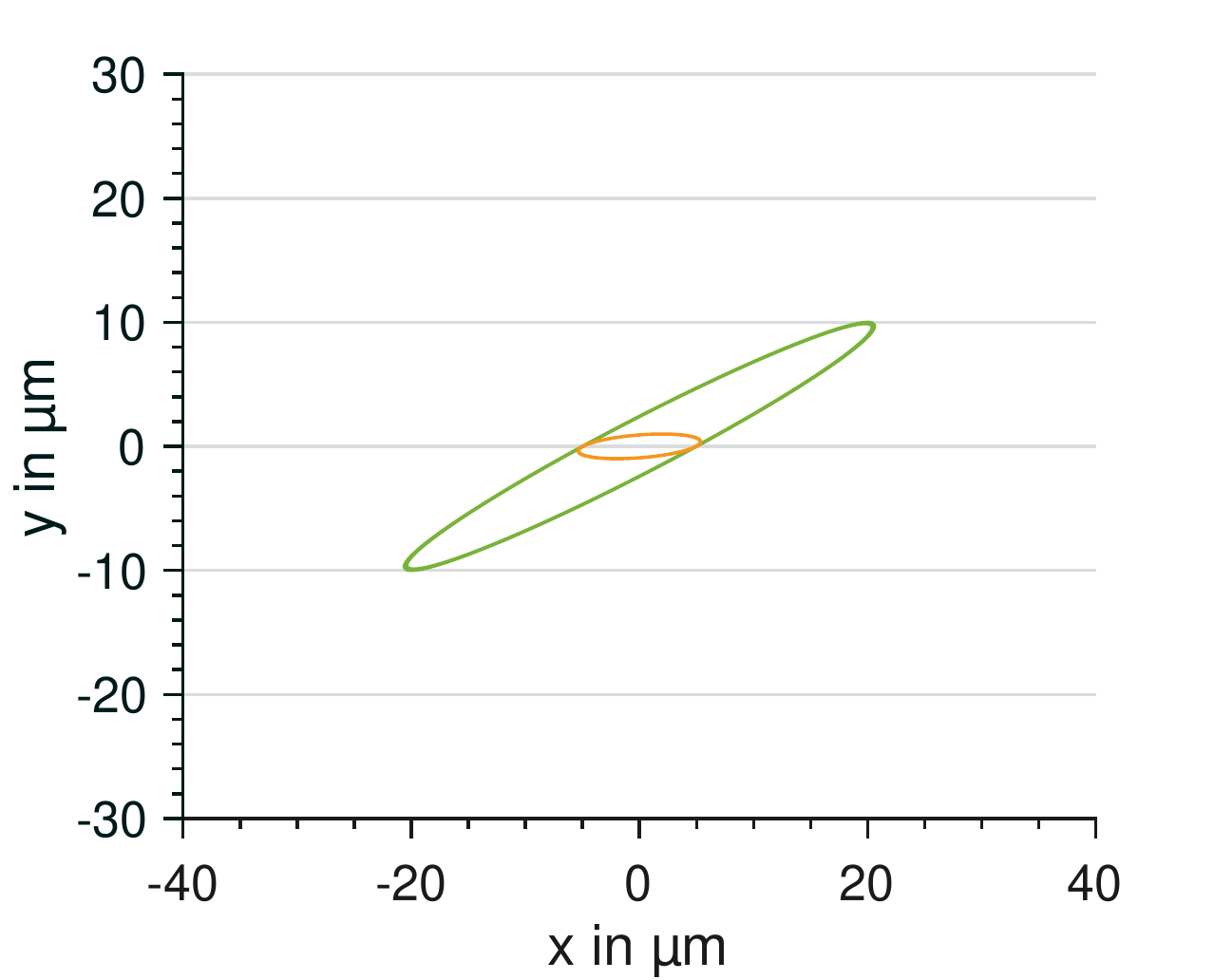}}
	\caption{(a) The optical potential along the axis of gravity is tilted, which reduces the effective trap depth in one direction. (b) DSMC simulation for an evaporative sequence for different gravitational accelerations. The temperature evolution marginally differs between the cases of microgravity (green) and grond based operation (orange). The efficiency of evaporation, and therefore the dimension of evaporation, are equal in both cases. (c) The trajectories of two particles in a realistic dipole trap potential with $z=0$ (z is the trapping beam direction). The direction of kinetic energy is transferred within a few oscillations. This effect increases towards the trap edges, where the potential strongly deviates from it's harmonic approximation. (d) The trajecetories of two particles in a harmonic potential. The direction of kinetic energy is conserved for all velocities.}
	\label{fig:Sim}     
\end{figure*}
\section{Evaporation}
\label{sec:Evap}
Evaporative cooling from an optical trap is carried out by a reduction of the trap depth $U_0$. As it is depicted in Figure~\ref{fig:Sim}(a), the potential is tilted along the direction of gravity. This leads to an effectively lowered trap depth $U_\text{eff}$ in one direction. Therefore, in the presence of gravity, evaporated atoms are preferably expelled along this direction, while there is no distinction in microgravity.\newline
An evaporation, where the possibility to leave the trap depends only on the kinetic energy in one direction ($E_g>U_0$), is called one dimensional~\cite{Ketterle1996}. Its efficiency is calculated to be reduced by a factor of $4\frac{U_0}{T}$~\cite{Surkov1996} in comparison to three dimensional evaporation. The reduction can easily be understood, since atoms have to collide with each other in order to produce atoms with sufficiently high kinetic energies to leave the trap. This is the limiting process in evaporation. If the probability to leave further depends on the atomic post-encounter direction, generally more collisions are needed.\newline  
We simulated the process of evaporation for different values of gravitational acceleration with the DSMC (direct simulation Monte Carlo) method~\cite{Bird1994a, Wu1996} and the results are shown in Figure~\ref{fig:Sim}(b). The difference in temperature evolution between \SI{1}{\gram} and \SI{0}{\gram} is found to be marginal. Therefore the dimension of evaporation has to be (almost) similar in both cases. This result is in good agreement with the measurement from a strongly tilted trap that was published by Hung \textit{et al.}~\cite{Hung2008}. The reason for this behavior becomes obvious by tracing the trajectories of atoms in a realistic dipole potential, as it is done in Figure~\ref{fig:Sim}(c). While the direction of kinetic energy is conserved in a harmonic confining potential (Figure~\ref{fig:Sim}(d)), it can be transferred to others in a realistic dipole potential. This is caused by the large anharmonicities at the trap edges. The figure shows two trajectories. One, for an atom closer to the trap center (orange line) and a second one for an atom that reaches further out. In the second situation, that more accurately represents the case of an evaporating atom, the transfer of kinetic energy between directions is strongly increased, the potentials are mixing. In this case, the evaporation becomes three dimensional, since the initial post-encounter direction does not affect the probability for an atom to be evaporated. \newline
This explanation is still in agreement with the decreased temperature evolution in the case of an increased gravitational acceleration of \SI{5}{\gram}. Here the decisive difference is the reduction of the effective trap depth, which gets large enough to cause an observable temperature decrease.\newline
These results were confirmed with an experimental evaporative sequence in the drop tower. The dipole trap power is lowered in four linear segments whith the respective final value, following $P=P_0\cdot e^{-t/\tau}+P_{f}$ where $P_0$ is the initial laser power of \SI{7}{\watt} (reduced by the offset), t is the time, $\tau$ represents the ramps time constant which was chosen to be \SI{700}{\milli\second} and $P_{f}$ is the final power for an infinite time. To investigate the evaporative process, the sequence was stopped after \SI{0.5}{\second}, \SI{1}{\second}, \SI{1.5}{\second} and \SI{2}{\second} respectively, followed by a free evolution time of up to 10 ms for temperature determination.\newline
\begin{figure}[tbp]
	\centering
	\includegraphics[width=0.35\textwidth]{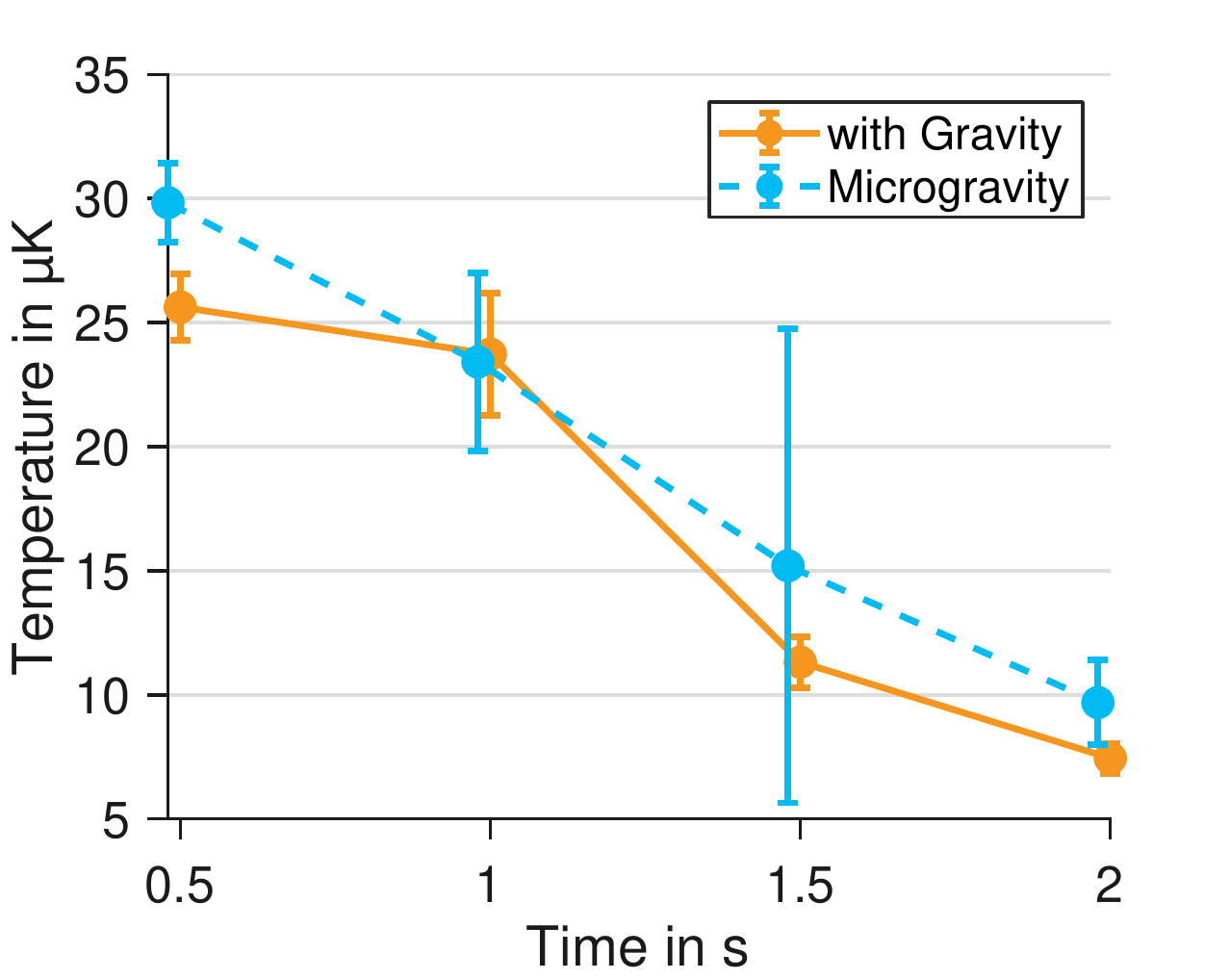}
	\caption{Experimentally realized evaporation on ground (blue) and in microgravity (orange). Apart form the increased temperature, caused by a reduced TDM performance in microgravity, both evaporation curves are equal within the measurement uncertainties. These suffer from a limited number of underlying statistics, since at most two experiments per day, for approximately two weeks (one measurement campaign), are available.}
	\label{fig:Evap1}   
\end{figure}
We characterized the initial trap frequencies in our system to be \SI{944}{\hertz} in the radial and \SI{9.2}{\hertz} in the axial direction by parametric heating and direct observation of oscillations, giving a trapping focus beam waist of \SI{45}{\micro\meter}. This leads to a trap depth of $U_0=k_B\times \SI{186}{\milli\kelvin}$ at the beginning of our evaporation sequence. \newline
Precise temperature determinations turned out to be challenging due to the low repetition rate of two drops per day. Since it was not possible to guarantee identical starting conditions for the evaporation over the course of weeks, the recorded data suffers from a large measurement uncertainty. Nevertheless, two important conclusions can be gathered from this dataset: \newline
First, the evaporation from an optical dipole trap works under microgravity. The slightly increased temperature evolution throughout the sequence is attributed to a worse TDM cooling in the rough drop tower conditions. Second, in combination with the simulation, one can assume the efficiency of evaporative cooling to be (almost) equal with and without the effect of gravity. \newline

\section{Conclusion}
\label{sec:Conc}
We were able to demonstrate the loading of, and evaporative cooling from an optical dipole trap in microgravity. The efficiency of evaporative cooling in weightlessness was calculated to be equal to ground based experiments. These findings were made with DSMC simulations and could be attributed to the anharmonicity of the confining potential. Furthermore, the theoretical results could be confirmed experimentally.  \newline
Our results prove this kind of experiments to be feasible in a space environment as well and could lead e.g. to improved atom interferometers, miscibility investigations and optical lattices in microgravity on time scales of several tens of seconds.\newline
The next goal is to demonstrate the creation of an all-optical BEC which is completely prepared in microgravity. Therefore, the trapping beam will be actively pointing stabilized giving the opportunity for a crossed beam configuration with decreased evaporation times.

\begin{acknowledgments}
This project is supported by the German Space Agency~(DLR) with funds provided by the Federal Ministry for Economic Affairs and Energy~(BMWi) due to an enactment of the German Bundestag under Grant No. DLR 50WM1641~(PRIMUS-III).
D.S. is grateful for personal funding by the Federal Ministry of Education and Research (BMBF) through the funding program Photonics Research Germany under contract number 13N14875.
\end{acknowledgments}

\bibliography{first_dipole_paper}

\end{document}